\documentclass[aip,pof,reprint,onecolumn]{revtex4-1} 
\usepackage{natbib}
\usepackage{amsbsy}
\usepackage{comment}
\usepackage{psfrag}
\usepackage{amsmath}
\usepackage{soul}
\usepackage{mathrsfs} 
\usepackage{amssymb}
\usepackage{mathtools}
\usepackage{mathptmx}      
\usepackage{booktabs}
\usepackage{graphicx}
\usepackage[usenames,dvipsnames,svgnames,table]{xcolor} 
\usepackage[colorlinks=true,
            linkcolor=blue,
            urlcolor=blue,
            citecolor=BrickRed]{hyperref}
            
\usepackage{mathrsfs}  

\usepackage{setspace}
\usepackage{makecell}

\usepackage{mathtools}
\usepackage{amssymb}
\usepackage{amsmath}
\usepackage[abs]{overpic}
\usepackage{xcolor,varwidth}            

\usepackage[utf8]{inputenc}
\usepackage[T1]{fontenc}
\usepackage{mathptmx}
\usepackage{CJKutf8}

\begin{document}
\begin{CJK*}{UTF8}{gbsn}
\title{Grid-point and time-step requirements for direct numerical simulation and large-eddy simulation}
\author{Xiang I. A. Yang (杨翔)}
\email{xzy48@psu.edu}
\affiliation{Department of Mechanical Engineering, Pennsylvania State University, State College, PA, 16802, USA}
\author{Kevin P. Griffin}
\affiliation{Center for Turbulence Research, Stanford University, Stanford, CA, 94305, USA}

\begin{abstract}
We revisit the grid-point requirement estimates in Choi and Moin [{\it Phys. Fluid}, {\bf 24}, 011702 (2012)] and establish more {general grid-point requirements for direct numerical simulations (DNS) and large-eddy simulations (LES) of a spatially developing turbulent boundary layer}.
We show that, by allowing the local grid spacing to scale with the local Kolmogorov length scale, the grid-point requirement for DNS of a spatially developing turbulent boundary layer is $N\sim Re_{L_x}^{2.05}$ rather than $N\sim Re_{L_x}^{2.64}$ as suggested by Choi and Moin, where $N$ is the number of grid points and $L_x$ is the length of the plate.
In addition to the grid-point requirement, we estimate the time-step requirement for DNS and LES.
We show that, for a {code that treats the convective term explicitly}, the time steps required to get converged statistics are $N_t\sim Re_{L_x}/Re_{x_0}^{6/7}$ for wall-modeled LES and $N_t\sim Re_{L_x}/Re_{x_0}^{1/7}$ for wall-resolved LES and DNS (with different prefactors), where $Re_{x_0}$ is the inlet Reynolds number.
The grid-point and time-step requirement estimates allow us to estimate the overall cost of DNS and LES.
According to present estimates, the costs of DNS, wall-resolved LES and wall-modeled LES scale as $Re_{L_x}^{2.91}$, $Re_{L_x}^{2.72}$, and $Re_{L_x}^{1.14}$. 
\end{abstract}

\maketitle
\end{CJK*}


\section{Introduction}

Many fields where computational fluid dynamics (CFD) plays an important role are cost-driven, and therefore, the grid-point and time-step requirements for CFD calculations are important practical issues.
{The first formal estimate for the grid-point requirement is by Chapman. \cite{chapman1979computational}
He estimated the grid-point requirement for wall-resolved large-eddy simulation (WRLES) of a spatially developing turbulent boundary layer and arrived at the estimate $N\sim Re_{L_x}^{9/5}$, where $N$ is the number of grid point, $Re_{L_x}=U_\infty L_x/\nu$ is the Reynolds number, $U_\infty$ is the freestream velocity, $L_x$ is the streamwise length of the flat plate, and $\nu$ is the kinematic viscosity.
Choi and Moin \cite{choi2012grid} revisited Chapman's estimate.
By invoking more accurate Reynolds number scalings for the skin friction coefficient and the boundary-layer thickness, Choi and Moin arrived at the estimate $N\sim Re_{L_x}^{2.65}$, $N\sim Re_{L_x}^{1.86}$, and $N\sim Re_{L_x}^1$ for direct numerical simulation (DNS), WRLES, and wall-modeled large-eddy simulation (WMLES), respectively.
Rezaeiravesh et al. \cite{Rezaeiravesh2016OnGR} generalized Choi and Moin's estimate by allowing for general Reynolds number scalings for the skin friction and the boundary layer thickness.
Liefvendahl and Fureby \cite{liefvendahl2017grid} followed Rezaeiravesh et al. and estimated the computational cost of ship hull hydrodynamics simulations.}
The objective of this paper is to revisit the grid-point requirements for DNS, WRLES, and WMLES of a spatially developing turbulent boundary layer and estimate the time-step requirement for DNS, WRLES, and WMLES.

\begin{figure}
    \centering
    \includegraphics[height=2.5in]{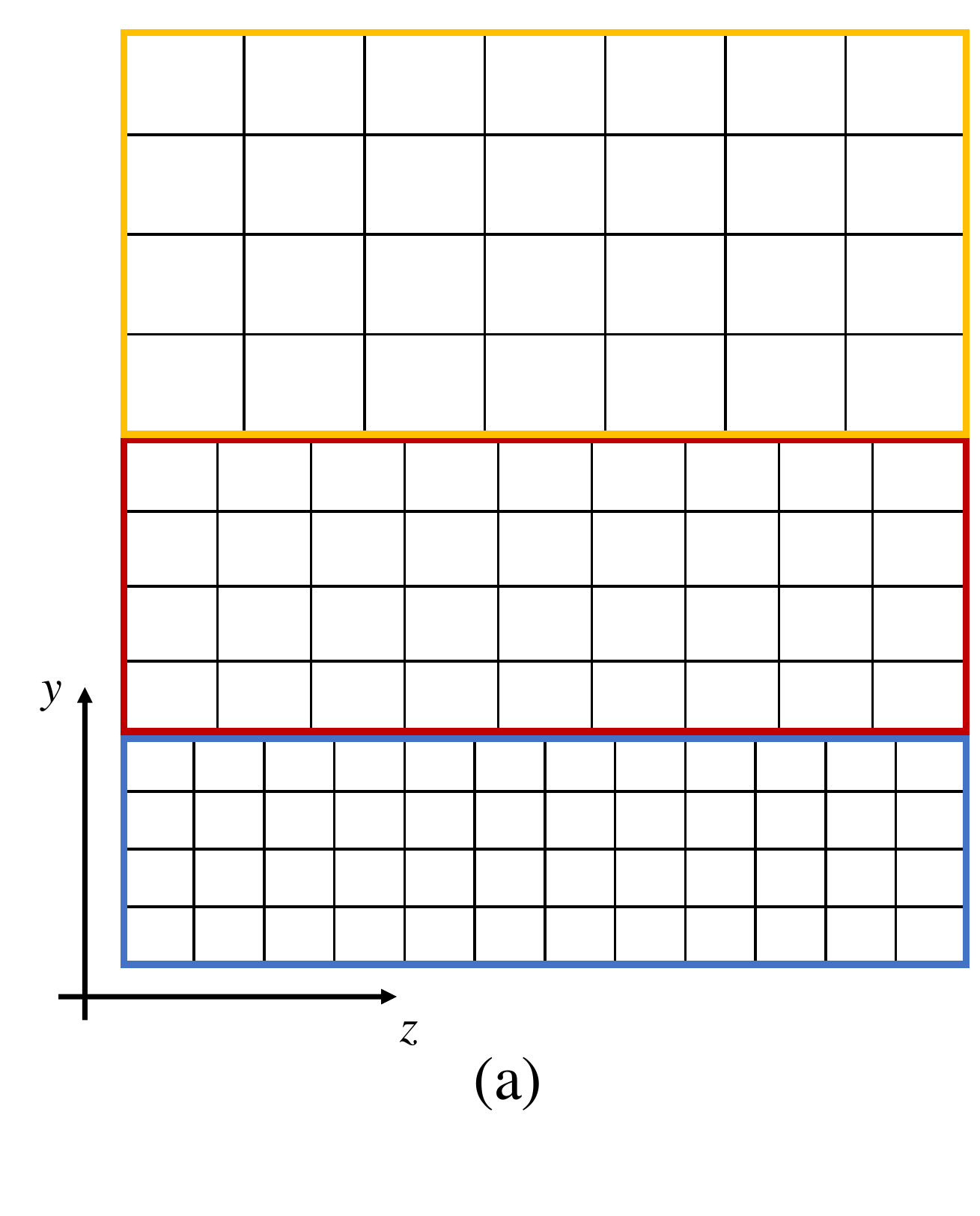}~~~~\includegraphics[height=2.5in]{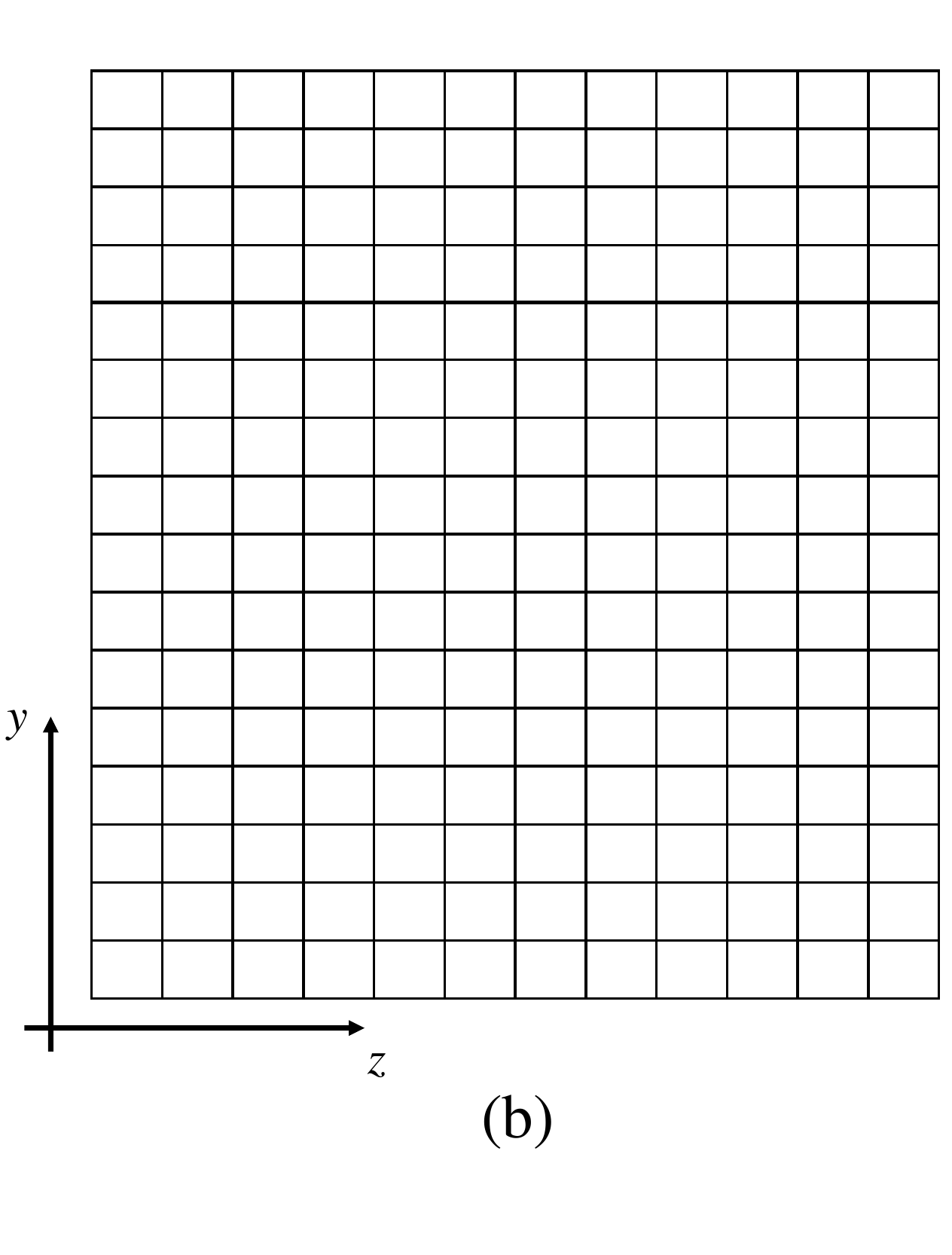}~~~~\includegraphics[height=2.5in]{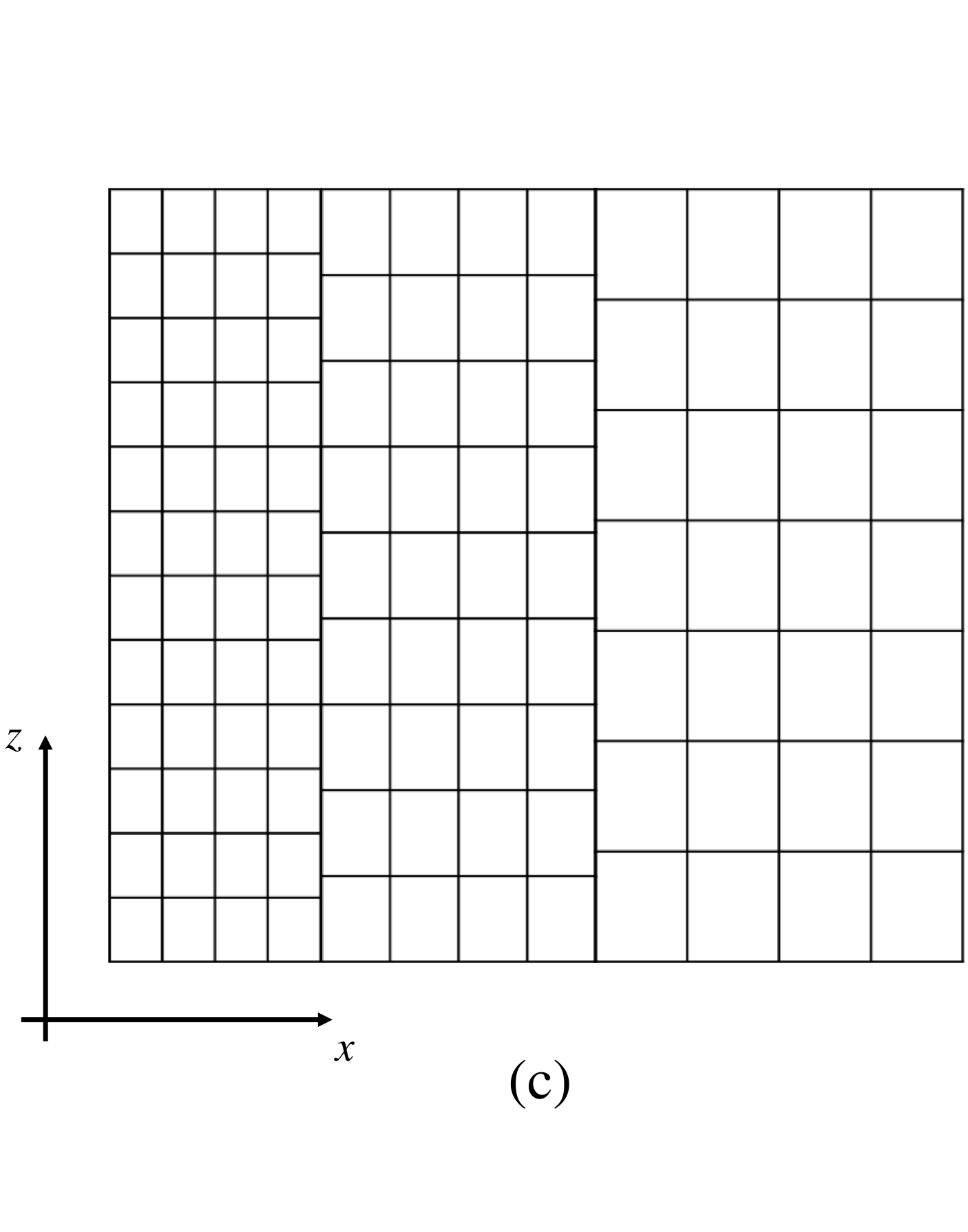}
    \caption{(a) A sketch of nested grids in a wall-normal/spanwise plane. Three blocks of grids are sketched. Each block contains four layers of grids. The spanwise (and streamwise) grid spacing coarsens from block to block.
    The first block contains $N_1=48$ grid points.
    The second block contains $N_2=36$ grid points.
    The third block contains $N_3=28$ grid points.
    (b) A sketch of an un-nested grid.
    (c) A sketch of the assumed streamwise/spanwise grid in Choi and Moin. \cite{choi2012grid}
    The grid spacing depends on $x$.
    Here $x$, $y$, $z$ are the streamwise, wall-normal, and the spanwise directions.
    }
    \label{fig:mesh}
\end{figure}

We begin our discussion by pointing out two weak points in the Chapman and Choi \& Moin's  estimates.
In arriving at his estimate, Chapman assumed the use of nested grids, as is sketched in figure \ref{fig:mesh} (a), where the streamwise, the wall-normal, and the spanwise grid spacings $\Delta x$, $\Delta y$, and $\Delta z$ double from block to block.
He then concluded that the total number of grid points in the boundary layer is $N=N_1[1+1/4+(1/4)^2+(1/4)^3+...]=4/3N_1$, giving rise to the coefficient 4/3: the total number of grid points $N$ is 4/3 the number of grid points in the wall layer $N_1$.
Choi and Moin followed Chapman and invoked the relation $N=4/3N_1$. \cite{choi2012grid}
However, the factor $N/N_1$, generally speaking, depends on the grid coarsening from one block to the next.
If the grid spacing coarsens by a factor of 1.5, $N/N_1$ would be 2, leading to arbitrariness in the grid-point estimates.
This arbitrariness {is} the first weak point, and we will try to remove {this weak point by allowing the grid spacing to depend on  physical length scales}.
The second weak point in Choi and Moin's analysis is the use of un-nested grids in the wall-normal-spanwise plane for the estimate of the grid-point requirement for DNS. \cite{choi2012grid}
Figure \ref{fig:mesh} (b) is a sketch of an un-nested grid: {the grid spacing is the same in the wall layer and the outer layer}.
In a wall-bounded flow, the Kolmogorov length scale $\eta$ is an increasing function of $y$, and therefore, the use of un-nested grids is overly conservative (as acknowledged by Choi and Moin).
Although Choi and Moin used un-nested grid in the spanwise-wall-normal plane, they did allow the grid spacing to depend on $x$, which necessarily leads to nested grids in the streamwise direction, as shown in figure \ref{fig:mesh} (c).
{In the present work, we will remove the second weak point by allowing nested grid in all three Cartesian directions.}

The rest of the paper is organized as follows.
In section \ref{sec:grid_pt_req} of this article, we resolve the aforementioned weaknesses by acknowledging:
first, the sizes of energy containing eddies in a turbulent boundary layer scale as their distances from the wall; \cite{marusic2019attached}
second, the Kolmogorov length scale is a function of the wall-normal distance.
We would follow Choi and Moin and invoke the following relations {for the boundary-layer thickness $\delta$ and the friction coefficient $c_f$}
\begin{equation} 
\begin{split}
    \frac{\delta}{x}=0.16Re_x^{-1/7},\\
    c_f=0.027Re_x^{-1/7},
\end{split}   
\label{eq:emp}
\end{equation}
which are more accurate {for high Reynolds number flows ($10^6 \leq Re_x \leq 10^9$)} than the ones used by Chapman. \cite{nagib2007approach, monkewitz2007self}
Throughout the article, CM is the abbreviation for Ref. \onlinecite{choi2012grid}.
In section \ref{sec:time_step}, we estimate the time-step requirements for DNS, WRLES, and WMLES. 
Concluding remarks are given in section \ref{sec:conclusion} following a brief discussion in section \ref{sect:discussion}. 
{We include a table of nomenclature in the appendix.}

\section{Grid-point requirement} \label{sec:grid_pt_req}

First, we determine the grid-point requirement for a WRLES.
While ensuring our derivation is self-contained, we will include references to CM and Chapman as ``check points'' for validation purposes.
These ``check points'' also help us to highlight the differences between our derivations and the ones in the previous work.
The local grid spacing in an LES should be proportional to the sizes of the local energy-containing eddies.
In a boundary layer and at a distance above the viscous sublayer, the sizes of the energy containing eddies scale as their distances from the wall.
Hence, the grid resolutions are such that $\Delta_x\sim y$, $\Delta_y\sim y$, $\Delta_z\sim y$, which gives rise to 
\begin{equation} 
    \Delta_x=\frac{1}{n_x}y,~~~\Delta_y=\frac{1}{n_y}y,~~{\rm and}~~~\Delta_z=\frac{1}{n_z}y,
    \label{eq:Delta}
\end{equation}
where $n_x\times n_y\times n_z$ is the number of grid points one deploys in a box of size $y\times y\times y$.
The typical values of $n_x$, $n_y$ and $n_z$ could be found in Ref \onlinecite{choi2012grid}, and we will not repeat these numbers here for brevity.
The number of grid points in a box $dx\times dy\times dz$ at a generic location $x$, $y$, $z$ is
\begin{equation} 
    dN_{wr}=\frac{dx}{\Delta_x}\frac{dy}{\Delta_y}\frac{dz}{\Delta_z}.
    \label{eq:dN1}
\end{equation}
Equation \eqref{eq:dN1} is valid in the wall layer and in the outer layer---thereby, we remove the first weak point in the analysis of CM.
Equation \eqref{eq:dN1} directly leads to
\begin{equation} 
\begin{split}
   N_{wr}=\int_V dN_{wr}=N(x<x_0)
   &+\int_{x_0}^{L_x} \frac{dx}{\Delta_{x,w}} \int_0^{L_z} \frac{dz}{\Delta_{z,w}}  ~\left[N_{y,30}\right] \\
   &+\int_{x_0}^{L_x} dx \int_0^{L_z} dz \int_{30\nu/u_\tau}^\delta dy ~\left[\frac{n_x}{y}\frac{n_y}{y}\frac{n_z}{y}\right],
\end{split}
\label{eq:N1}
\end{equation}
where $V$ is the fluid region, $N(x<x_0)$ is the number of grid points one would need for $x<x_0$, $N_{y,30}$ is the number of $y$ grid points one would need from the wall to $y^+=30$, $\Delta_{x,w}$ and $\Delta_{z,w}$ are the grid spacings in the streamwise and the spanwise directions for $y^+<30$.
To highlight and clarify the variables that are being integrated with respect to, we write first the differentials and their integration limits and second the function to be integrated (a common practice in today's physics/math literature).
The second term in Equation \eqref{eq:N1} is practically equation 12 in CM, but rather than $y^+=30$, Choi and Moin used $l_y^+$.
Following CM, we have
\begin{equation} 
\begin{split}
    \int_{x_0}^{L_x} \frac{dx}{\Delta_{x,w}} \int_0^{L_z} \frac{dz}{\Delta_{z,w}}  ~\left[N_{y,30}\right]&=L_zN_{y,30}\frac{1}{\Delta_{x,w}^+}\frac{1}{\Delta_{z,w}^+}\int_{x_0}^{L_x} \frac{u_\tau^2}{\nu^2} dx\\ &=0.0158\frac{L_z}{L_x}\frac{N_{y,30}}{\Delta_{x,w}^+\Delta_{z,w}^+}Re_{L_x}^{13/7}\left[1-\left(\frac{Re_{x_0}}{Re_{L_x}}\right)^{6/7}\right]\\
    &=0.0158\frac{L_z}{L_x}\frac{N_{y,30}}{\Delta_{x,w}^+\Delta_{z,w}^+}\left[Re_{L_x}^{13/7}+O(Re_{L_x})\right].
    \label{eq:N12}
\end{split}
\end{equation}
{The result in Equation \eqref{eq:N12} is consistent with equation 13 in CM.}
This serves as a validation of our derivation.
The third term in Equation \eqref{eq:N1} is
\begin{equation}
\begin{split}
    \int_{x_0}^{L_x} dx &\int_0^{L_z} dz \int_{30\nu/u_\tau}^\delta dy ~\left[\frac{n_x}{y}\frac{n_y}{y}\frac{n_z}{y}\right]=L_zn_xn_yn_z\int_{x_0}^{L_x}\frac{1}{2}\left(\frac{u_\tau^2}{900\nu^2}-\frac{1}{\delta^2}\right)dx\\
    &=C_1\frac{L_z}{L_x}n_xn_yn_z\left\{ Re_{L_x}^{13/7}\left[1-\left(\frac{Re_{x_0}}{Re_{L_x}}\right)^{6/7}\right]+C_2Re_{L_x}^{2/7}\left[1-\left(\frac{Re_{L_x}}{Re_{x_0}}\right)^{5/7}\right]\right\}\\
    &= C_1\frac{L_z}{L_x}n_xn_yn_z \left[Re_{L_x}^{13/7}+O(Re_{L_x})\right],
    \label{eq:N13}
\end{split}
\end{equation}
where $C_1=8.75\times 10^{-6}$ and $C_2=27.3/C_1$.
In practice, $N_{y,30}\approx n_y$. 
By definition, $n_x={30}/{\Delta_{x,w}^+}$ and $n_z={30}/{\Delta_{z,w}^+}$.
Plugging Equations \eqref{eq:N12} and \eqref{eq:N13} into Equation \eqref{eq:N1}, we have
\begin{equation} 
\begin{split}
    N_{wr}&=\left(0.0158+8.75\times 10^{-6}\times 30^2\right)\frac{L_z}{L_x}\frac{n_{y}}{\Delta_{x,w}^+\Delta_{z,w}^+}Re_{L_x}^{13/7}+O(Re_{L_x})\\
    &=0.024\frac{L_z}{L_x}\frac{n_{y}}{\Delta_{x,w}^+\Delta_{z,w}^+}Re_{L_x}^{13/7}+O(Re_{L_x}).
\end{split}   
\label{eq:NWRLES}
\end{equation}
The estimate turns out to be only slightly different from that in CM---with the same Reynolds number scaling and with the pre-factor being 0.021 in CM and 0.024 here. 
{Again, the use of nested grids is assumed.
}

\begin{figure}
    \centering
    \includegraphics[width=0.4\textwidth]{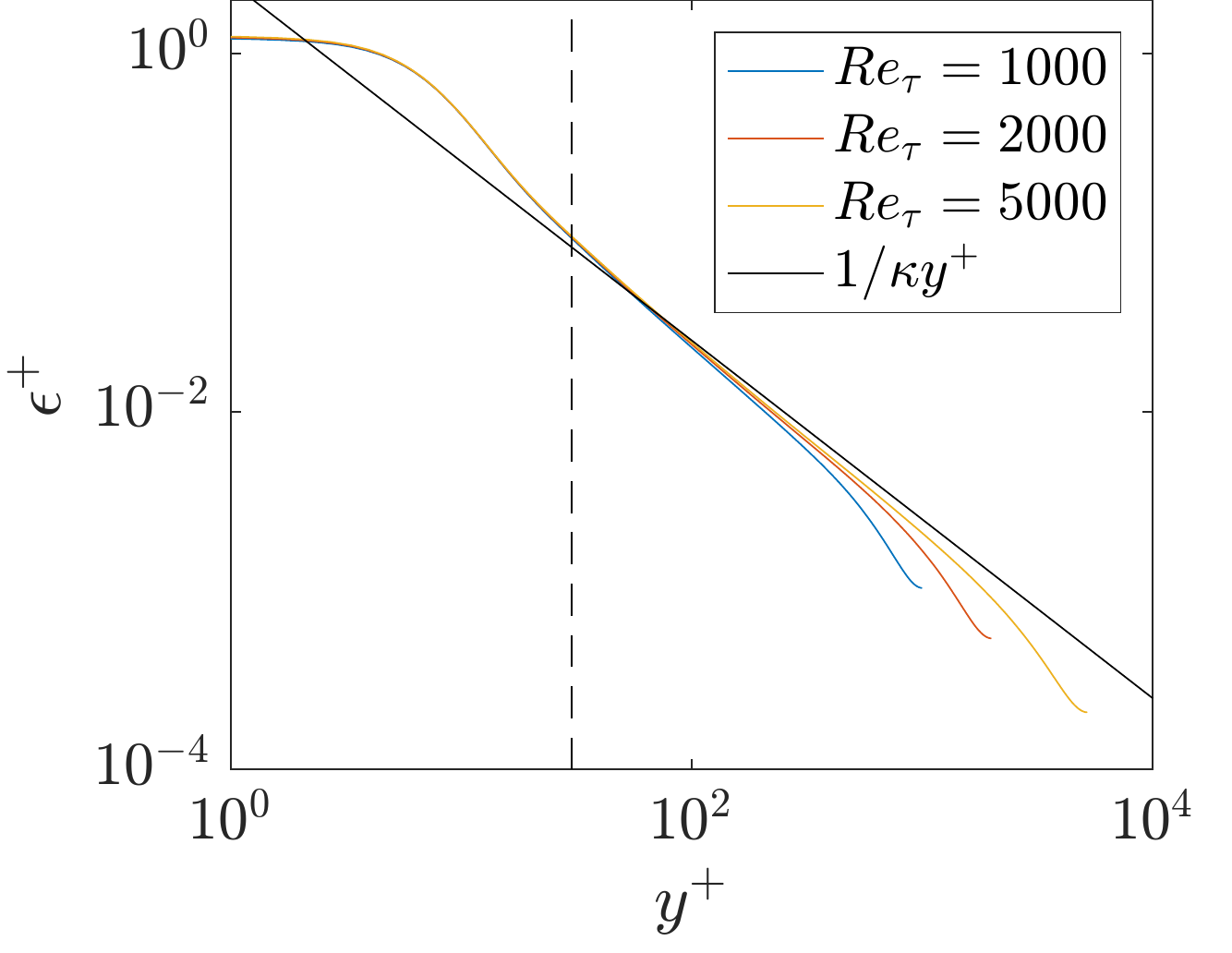}
    \caption{Dissipation rate in channel flow at $Re_\tau=1000$, 2000, and 5200.
    Details of the data could be found in Refs. \onlinecite{graham2016web,hoyas2006scaling,lee2015direct}.
    Here, $\epsilon$ is the total dissipation, and it includes $\nu (dU/dy)^2$.
    The superscript $+$ denotes normalization by wall units.
    }
    \label{fig:eps}
\end{figure}

Second, we determine the grid-point requirement for DNS.
It follows from Equation \eqref{eq:emp} that
\begin{equation} 
    Re_\tau=0.019Re_{x}^{11/14},
\end{equation}
for $10^6<Re_x<10^9$, or $9.8\times 10^2<Re_\tau<2.2\times 10^5$.
For flows at these high Reynolds numbers, 
\begin{equation} 
    {\rm Dissipation}=\epsilon\approx {\rm Production}={-}\left<u'v'\right>\frac{dU}{dy}{\approx} \frac{u_\tau^3}{\kappa}\frac{1}{y}
    \label{eq:eps}
\end{equation}
is a fairly good working approximation of the dissipation in the logarithmic layer, i.e., for $y^+>30$, where $\kappa=0.4$ is the von Karman constant.
Figure \ref{fig:eps} compares Equation \eqref{eq:eps} and data at $Re_\tau=1000$, $2000$, and $5200$.
It follows from Equation \eqref{eq:eps} that the Kolmogorov length scale is
\begin{equation} 
    \eta=\left(\frac{\nu^3}{\epsilon}\right)^{1/4} \sim \left(\frac{\nu}{u_\tau}\right)^{3/4}y^{1/4},
\end{equation}
for $y^+>30$, where $\epsilon$ is the total dissipation.
For $y^+<30$, a conservative estimate of the dissipation rate is $\epsilon^+=1$, \cite{choi2012grid} and a conservative estimate of the Kolmogorov length scale follows: $\eta^+=1$.
The grid resolution in a DNS is proportional to the local Kolmogorov length scale, i.e.,
\begin{equation} 
    \Delta_x=C_x\eta,~~~\Delta_y=C_y\eta,~~{\rm and}~~~\Delta_{{z}}=C_z\eta,
\end{equation}
where $C_x$, $C_y$, and $C_z$ are O(1) constants.
Hence, the number of grid points needed for a DNS of a spatially developing boundary layer is
\begin{equation} 
\begin{split}
    N_{DNS}=\int_V \frac{dx}{\Delta_x}\frac{dy}{\Delta_y}\frac{dz}{\Delta_z}=&N(x<x_0)
   +\int_{x_0}^{L_x} \frac{dx^+}{C_x} \int_0^{L_z} \frac{dz^+}{C_z}  ~\frac{30}{C_y}\\
   &+\int_{x_0}^{L_x} dx \int_0^{L_z} dz \int_{30\nu/u_\tau}^\delta dy ~\left[\frac{1}{C_x C_yC_z}\left(\frac{\nu}{u_\tau}\right)^{-9/4}\frac{1}{y^{3/4}}\right].
\end{split}
\label{eq:N2} 
\end{equation}
A 9/4 exponent emerges as expected.
This 9/4 gives rise to the $N_{DNS}\sim Re^{9/4}$ in Ref \onlinecite{rogallo1984numerical}.
The second term in Equation \eqref{eq:N2} is
\begin{equation} 
\begin{split}
\int_{x_0}^{L_x} \frac{dx^+}{C_x} \int_0^{L_z} \frac{dz^+}{C_z}  ~\frac{30}{C_y}&=L_z\frac{1}{C_x C_y C_z}\frac{30}{\nu^2}\int_{x_0}^{L_x} u_\tau^2 dx\\
&=0.473\frac{L_z}{L_x}\frac{1}{C_xC_yC_z}Re_{L_x}^{13/7}\left[1-\left(\frac{Re_{x_0}}{Re_{L_x}}\right)^{6/7}\right]\\
&=0.473\frac{L_z}{L_x}\frac{1}{C_xC_yC_z}Re_{L_x}^{13/7}+O(Re_{L_x}).
\end{split}
\label{eq:N21}
\end{equation}
The third term in Equation \eqref{eq:N2} is
\begin{equation} 
\begin{split}
    &~~~\int_{x_0}^{L_x} dx \int_0^{L_z} dz \int_{30\nu/u_\tau}^\delta dy ~\left[\frac{1}{C_x C_yC_z}\left(\frac{\nu}{u_\tau}\right)^{-9/4}\frac{1}{y^{3/4}}\right]\\
    &=\frac{1}{C_xC_yC_z}\frac{1}{\nu^{9/4}}L_z\int_{x_0}^{L_x}4 u_\tau^{9/4}\left[\delta^{1/4}-(30\nu/u_\tau)^{1/4}\right]dx\\
    &=\frac{0.019}{C_xC_yC_z}\frac{L_z}{L_x}Re_{L_x}^{115/56}\left[1-\left(\frac{Re_{x_0}}{Re_{L_x}}\right)^{59/56}\right]-\frac{0.15}{C_xC_yC_z}\frac{L_z}{L_x}Re_{L_x}^{13/7}\left[1-\left(\frac{Re_{x_0}}{Re_{L_x}}\right)^{6/7}\right]\\
    &=0.019\frac{1}{C_xC_yC_z}\frac{L_z}{L_x}Re_{L_x}^{115/56}+O\left(Re_{L_x}^{13/7}\right).
\end{split}
\label{eq:N22}
\end{equation}
Plugging Equations \eqref{eq:N21} and \eqref{eq:N22} into Equation \eqref{eq:N2}, we have
\begin{equation} 
    N_{DNS}=0.019\frac{1}{C_xC_yC_z}\frac{L_z}{L_x}Re_{L_x}^{115/56}+O\left(Re_{L_x}^{13/7}\right).
    \label{eq:NDNS}
\end{equation}
Compared to the estimate in CM, i.e., $N_{DNS}\sim Re_{L_x}^{37/14}$, the estimate in Equation \eqref{eq:NDNS} indicates a much weaker dependence on $Re_{L_x}$ due to the use of a nested grid.
In arriving at Equation \eqref{eq:NDNS}, we assume $\epsilon\sim 1/y$ above $y^+=30$.
As shown in figure \ref{fig:eps}, this overestimates the dissipation rate in the wake layer, which in turn leads to an underestimate of the Kolmogorov length scale and a conservative estimate of the grid-point requirement.

{Next, we compare the grid-point requirements of WMLES, WRLES, and DNS.
By assuming that the grid spacing scales as the local boundary layer height $\delta$ and that the LES/wall-model matching location is at some fraction of $\delta$ (most wall models should have no difficulty complying with this requirement \cite{kawai2012wall,park2014improved,bose2014dynamic,yang2015integral,yang2018semi,yang2019predictive,huang2019wall,iyer2019analysis,bae2019dynamic,meneveau2020note,griffin2020new}), Choi and Moin concluded that $N_{wm}\sim Re_{L_x}^{1}$.}
In all, to resolve a spatially developing turbulent boundary layer, a WMLES requires $N_{wm}\sim Re_{L_x}^{1.00}$, a WRLES requires $N_{wr}\sim Re_{L_x}^{13/7}=Re_{L_x}^{1.86}$, and a DNS requires $N_{DNS}\sim Re_{L_x}^{115/56}=Re_{L_x}^{2.05}$.

\begin{figure}
    \centering
    \includegraphics[width=0.4\textwidth]{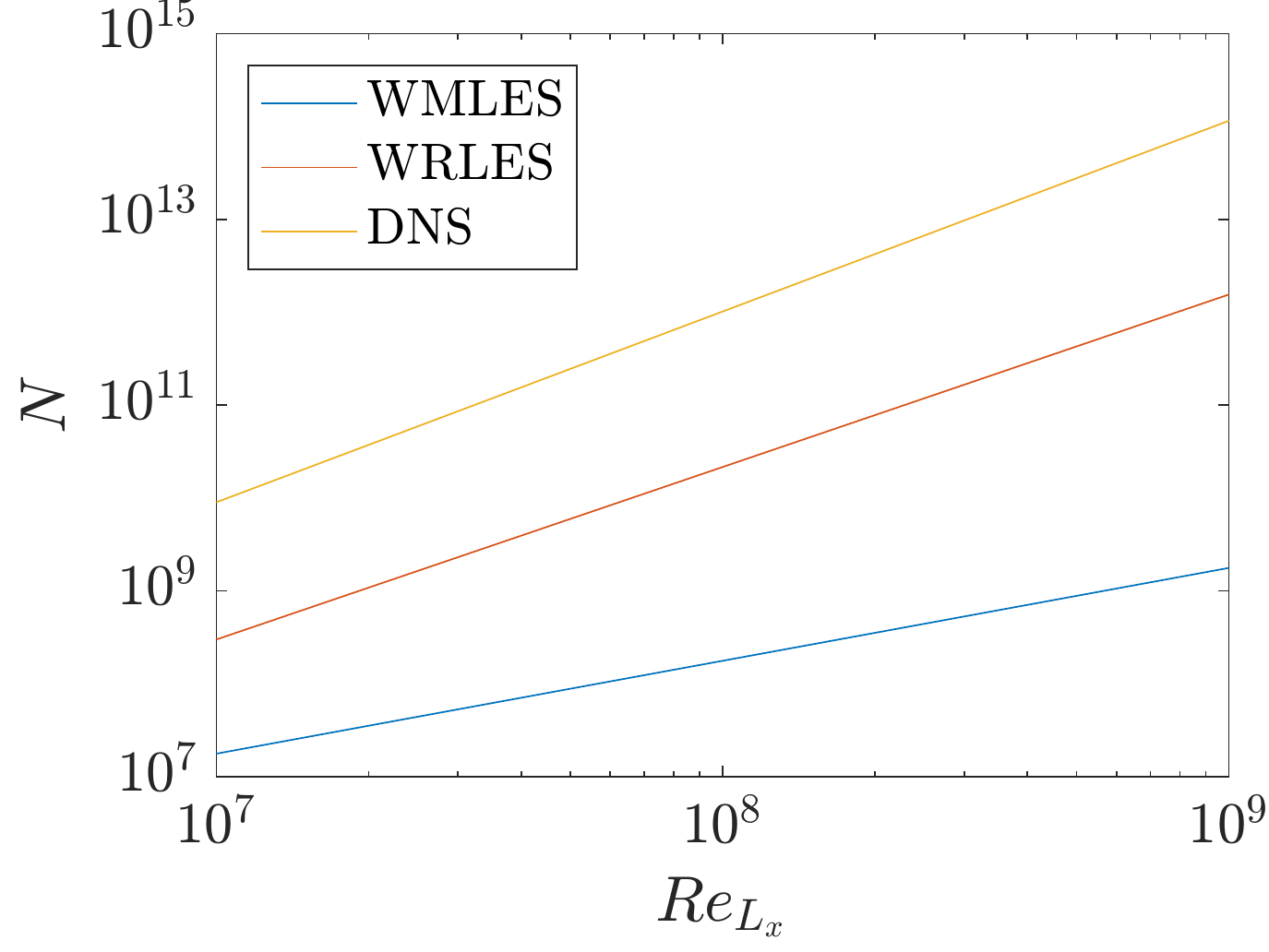}
    \caption{Number of grid points needed for DNS, WRLES, and WMLES of a spatially developing turbulent boundary layer.
    }
    \label{fig:N}
\end{figure}

Figure \ref{fig:N} shows the number of grid points for WMLES, WRLES, and DNS of a spatially developing turbulent boundary layer as a function of $Re_{L_x}$.
The grid-point requirement for WMLES is 
\begin{equation} 
    N_{wm}=54.7\frac{L_z}{L_x}n_xn_yn_zRe_{L_x}^{2/7}\left[\left(\frac{Re_{L_x}}{Re_{x_0}}\right)^{5/7}-1\right],
\end{equation}
following Ref \onlinecite{choi2012grid}.
When generating this figure, we have followed Choi and Moin and assume $Re_{x_0}=5\times 10^5$, $n_xn_yn_z=2500$, $n_y/\Delta_{w,x}^+\Delta_{w,z}^+=1/200$, and $L_x/L_z=4$.
The constants $C_xC_yC_z=125$.
We neglect the grid points needed for $x<x_0$ and retained only the leading order term.
As a result, the estimates are only valid for $Re_{L_x}\gg Re_{x_0}$.
According to figure \ref{fig:N}, in this Reynolds number range, DNS is about 100 times more costly than WRLES.
If the computing ability doubles every two years, i.e., if Moore's (or Koomey's\cite{koomey2010implications}) law holds, the Reynolds number range that is accessible to WRLES today will be accessible to DNS in about 13 years.
In addition, compared with WRLES, WMLES becomes increasingly more cost-efficient as the Reynolds number increases. \cite{larsson2016large,bose2018wall}

\section{Time-step requirement}
\label{sec:time_step}

The cost of a simulation is a direct function of the time-step requirement.
{In this section, we estimate the number of time steps required to get converged statistics.
Mathematically, 
\begin{equation}
    \text{Time step requirement}=\frac{\text{Physical time needed for converged statistics}}{\text{Time step size}}.
\end{equation}
{We first estimate the time step size, $dt$.
If one treats both the viscous term and the convective term implicitly, there is no stability requirement for the time step size.
A more common practice is to treat the viscous term implicitly and the convective term explicitly.\cite{mittal2008versatile,scalo2015compressible,bermejo2014confinement}
For these semi-implicit codes, the Courant–Friedrichs–Lewy (CFL) number has to be smaller than some $\mathcal{O}(1)$ value.}
For an incompressible flow solver, an estimate of the local CFL number is
\begin{equation} 
\begin{split}
    {\rm CFL}=\frac{U dt}{\Delta_x} &=\frac{u_\tau^2 dt}{\nu}\frac{U^+}{\delta^+/n_x}, ~~~{\rm for~WMLES};\\
    &=\frac{u_\tau^2 dt}{\nu}~\frac{U^+}{y^+/n_x}, ~~~{\rm for~WRLES};\\
    &=\frac{u_\tau^2 dt}{\nu}~\frac{U^+}{C_x\eta^+}, ~~~~{\rm for~DNS}, \\
\end{split}
\label{eq:CFL1}
\end{equation}
where we neglect fluctuations in the velocity (the root mean square of $U$ is small compared to $U$ itself).
For the purpose of this discussion, we require CFL$\leq 1$, and it follows from Equation \eqref{eq:CFL1} that
\begin{equation} 
\begin{split}
    dt\leq &\min_{x,y}\left[\frac{\nu}{u_\tau^2}\frac{\delta^+/n_x}{U^+}\right],~~{\rm for~WMLES};\\
    dt\leq&\min_{x,y}\left[\frac{\nu}{u_\tau^2}\frac{y^+/n_x}{U^+}\right],~~{\rm for~WRLES};\\
    dt\leq&\min_{x,y}\left[\frac{\nu}{u_\tau^2}\frac{C_x \eta^+}{U^+}\right],~~{\rm for~DNS},\\
\end{split}
\label{eq:dt2}
\end{equation}
where $U^+$ and $\eta^+$ are functions of $y^+$; $u_\tau$ and $\delta$ are functions of $x$; $n_x$ and $C_x$ are constants.
Figure \ref{fig:dt} shows $y^+/U^+$ and $\eta^+/U^+$ as a function of $y^+$ for channel flow at $Re_\tau=1000$, $2000$, and $5200$.
According to figure \ref{fig:dt}, $y^+/U^+$ attains its minimum, i.e., 1, at $y^+=1$, and $\eta^+/U^+$ attains its minimum, i.e., 0.13, at $y^+\approx 15$.
\begin{figure}
    \centering
    \includegraphics[width=0.6\textwidth]{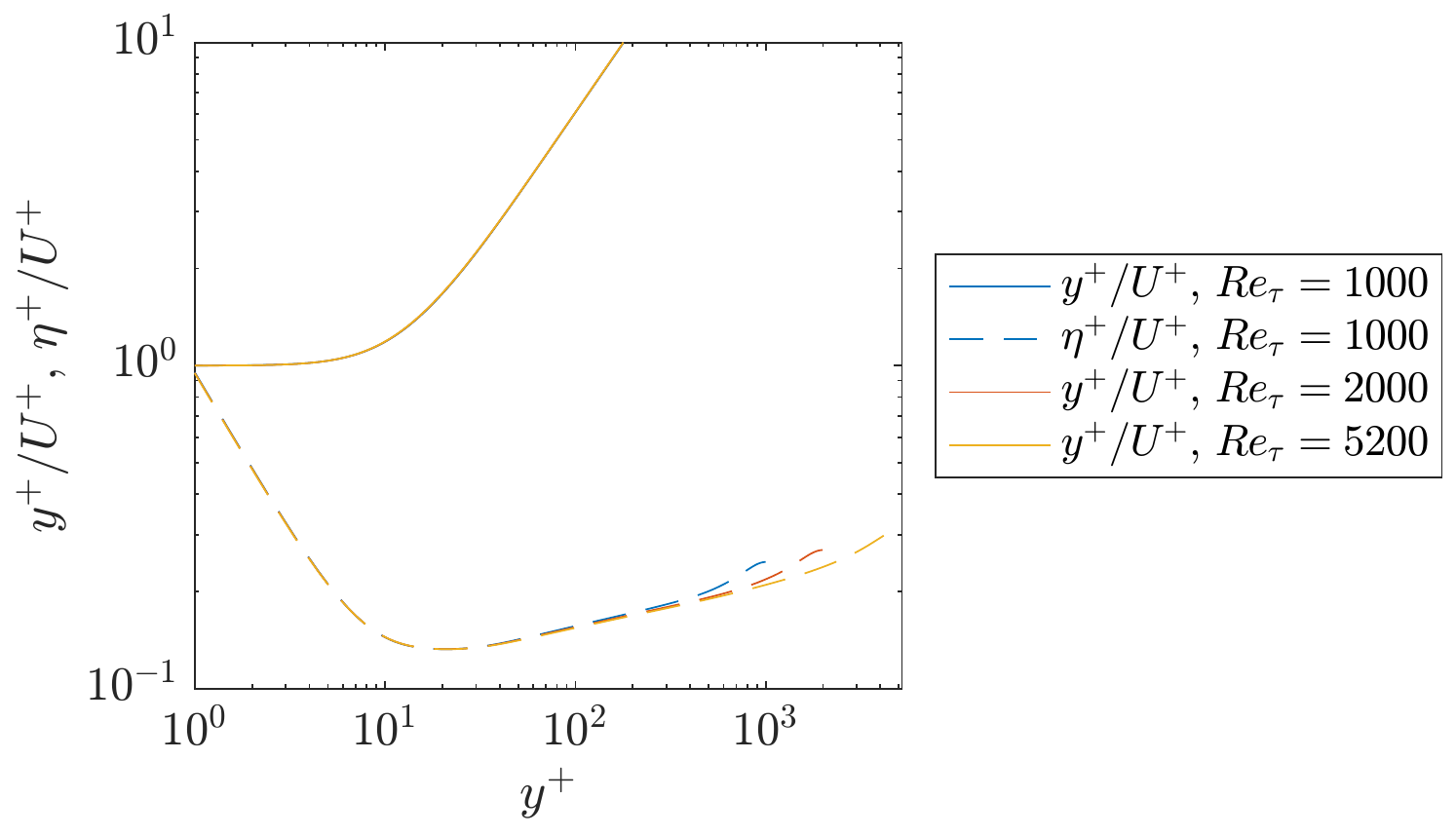}
    \caption{$y^+/U^+$ and $\eta^+/U^+$ as a function of the wall-normal distance in a channel at Reynolds numbers $Re_\tau=1000$, $2000$, and $5200$.
    Solid lines are for $y^+/U^+$.
    Dashed lines are for $\eta^+/U^+$.
    Different Reynolds numbers are color-coded, with blue for $Re_\tau=1000$, red for $Re_\tau=2000$, yellow for $Re_\tau=5200$.
    }
    \label{fig:dt}
\end{figure}
Hence, Equation \eqref{eq:dt2} reduces to
\begin{equation} 
\begin{split}
    dt\leq &\min_{x}\left[\frac{\nu}{u_\tau^2}\frac{\delta^+/n_x}{U_{\infty}^+}\right],~~{\rm for~WMLES};\\
    dt\leq&\min_{x}\left[\frac{\nu}{n_x u_\tau^2}\right],~~~~~~~~{\rm for~WRLES};\\
    dt\leq&\min_{x}\left[0.13\frac{C_x\nu}{u_\tau^2}\right],~~~~{\rm for~DNS}.\\
\end{split}
\label{eq:dt3}
\end{equation}
Plugging Equation \eqref{eq:emp} into Equation \eqref{eq:dt3} leads to
\begin{equation}
\begin{split}
    \frac{U_\infty^2dt}{\nu}\leq &\min_{x}\left[\frac{Re_x^{6/7}}{n_x}\right]=\frac{Re_{x_0}^{6/7}}{n_x},~~~~~~~~~~~~~~~~{\rm for~WMLES};\\
    \frac{U_\infty^2dt}{\nu}\leq&\min_{x}\left[\frac{74}{n_x}Re_x^{1/7}\right]=\frac{74}{n_x}Re_{x_0}^{1/7},~~~~~~~~~~{\rm for~WRLES};\\
    \frac{U_\infty^2dt}{\nu}\leq&\min_{x}\left[9.6C_xRe_x^{1/7}\right]=9.6C_xRe_{x_0}^{1/7},~~~{\rm for~DNS};\\
\end{split}
\label{eq:dt4}
\end{equation}
assuming that the inlet is at $x=x_0$.
Near the wall, i.e., at the $y$ locations where $y^+/U^+$ and $\eta^+/U^+$ attain their minimums, typical values are $C_x=1$, $n_x=1$, and Equation \eqref{eq:dt4} leads to
\begin{equation} 
\begin{split}
    \frac{U_\infty^2dt}{\nu}&\leq 0.2{Re_{x_0}^{6/7}},~~~{\rm for~WMLES};\\
    \frac{U_\infty^2dt}{\nu}&\leq 74Re_{x_0}^{1/7},~~~{\rm for~WRLES};\\
    \frac{U_\infty^2dt}{\nu}&\leq 9.6Re_{x_0}^{1/7},~~~{\rm for~DNS}.\\
\end{split}
\label{eq:dt5}
\end{equation}

Next, we estimate the physical time needed to get converged statistics. 
There are two important simulation time scales: the flow-through time $T_{ft} = L_x/U_\infty$ and the eddy-turn-over time $T_{eto} = \delta/u_\tau$. 
The flow-through time is associated with the transients of the simulation and the eddy-turn-over time is associated with statistical convergence of turbulence quantities. 
In order to get converged statistics, one needs to integrate for several (the exact number ranges from less than one hundred to more than two thousand \cite{lund1998generation, leonardi2010channel}) eddy-turnover times. 
Using Equation \eqref{eq:emp},
\begin{equation}
    T_{{eto}} = \frac{\delta}{u_\tau}|_{x=L_x} = 1.38 Re_{L_x}^{13/14} \frac{\nu}{U_\infty^2},
\end{equation}
and 
\begin{equation}
    T_{ft} = \frac{L_x}{U_\infty} = Re_{L_x} \frac{\nu}{U_\infty^2}.
\end{equation}
The flow-through time is slightly larger than the eddy-turn-over time for large $Re_{L_x}$. 
We will be conservative and assume that a constant number $C_t$ of flow-through times $T_{ft}$ must be integrated. 
This implies
\begin{equation} 
    N_t =C_t~{T_{ft}}/{dt} =C_t~ {Re_{L_x}}\Big/\frac{U_\infty^2 dt}{\nu}. 
    \label{eq:nt}
\end{equation}
It follows from Equations \eqref{eq:dt5} and \eqref{eq:nt} that
\begin{equation} 
\begin{split}
    N_t&=C_t\frac{Re_{L_x}}{0.2Re_{x_0}^{6/7}},~~~{\rm for~WMLES};\\
    N_t&=C_t\frac{Re_{L_x}}{74 Re_{x_0}^{1/7}},~~~{\rm for~WRLES};\\
    N_t&=C_t\frac{Re_{L_x}}{9.6Re_{x_0}^{1/7}},~~~{\rm for~DNS}.\\
    \label{eq:nt2}
\end{split}
\end{equation}

Equation \eqref{eq:nt2} is the time-step requirement for DNS, WRLES, and WMLES.
Nonetheless, it would be more instructive if we can write these time step estimates in terms of a single Reynolds number.
To do that, we need to estimate $L_x/x_0$. 
We consider two common applications. 
First, in a canonical simulation of the turbulent section of a boundary layer, the domain is usually constructed such that $L_x-x_0 = C_\delta \delta_{L_x}$, where $C_\delta$ is a constant (e.g. $C_\delta = 24$ was used in Ref \onlinecite{lund1998generation}). This choice combined with the correlations in Equation \eqref{eq:emp} imply
\begin{equation}
    \frac{x_0}{L_x} = 1 - 0.16 C_\delta Re_{L_x}^{-1/7}.
\end{equation}
In the high-Reynolds-number limit, $x_0/L_x \rightarrow 1$. Second, we consider the simulation of an airfoil where the transition location is geometrically imposed. Similar to the first case, this implies $x_0/L_x$ is approximately constant. For both of these cases, Equation \eqref{eq:nt2} becomes
\begin{equation} 
\begin{split}
    N_t&\sim Re_{L_x}^{1/7},~~~{\rm for~WMLES};\\
    N_t&\sim Re_{L_x}^{6/7},~~~{\rm for~WRLES};\\
    N_t&\sim Re_{L_x}^{6/7},~~~{\rm for~DNS}.\\
    \label{eq:nt3}
\end{split}
\end{equation}

\section{Discussion}
\label{sect:discussion}

Reynolds \cite{reynolds1990potential} proposed a back-of-the-envelope estimate for the time-step requirement.
He assumed a constant velocity in the computational domain and argued that the number of time steps $N_t$ scales as $N_t\sim N^{1/3}$, which leads to $N_t \sim Re_{L_x}^{115/168}$ for DNS, $N_t\sim Re_{L_x}^{13/21}$ for WRLES, and $N_t\sim Re_{L_x}^{1/3}$ for WMLES. 
These estimates for DNS ($N_t \sim Re_{L_x}^{0.68}$) and WRLES ($N_t \sim Re_{L_x}^{0.62}$) are slightly more optimistic than our predictions in Equation \eqref{eq:nt3} ($N_t \sim Re_{L_x}^{0.86}$). Whereas for WMLES, the estimate of Reynolds ($N_t \sim Re_{L_x}^{0.33}$) is more pessimistic than our prediction ($N_t \sim Re_{L_x}^{0.14}$). 

A crude estimate for the overall cost of a simulation on a single CPU scales with the number of grid points times the number of time steps, i.e., $N N_t$. 
However, provided access to very large computational resources and ideal software parallelization, the wall-clock time required for a simulation scales with $N_t$ only. 
For this reason, Equation \eqref{eq:nt3} provides a very optimistic outlook for WMLES in the age of exascale computing.

In section \ref{sec:time_step}, we have considered only the CFL condition.
If one treats the viscous term explicitly, he/she must account for the viscous stability condition:
\begin{equation} 
    {\rm viscous~stability~condition}=\frac{\nu dt}{\Delta^2} = \frac{u_\tau^2 dt}{\nu}~\frac{1}{ \Delta^{+,2}}\leq 1
    \label{eq:DCFL}
\end{equation}
where $\Delta$ is the grid spacing.
If a nested grid like the one in figure \ref{fig:mesh} (a) is used, the grid is locally isotropic, and therefore one could use $\Delta$ in any of the three directions for an estimate of the viscous stability condition.
For typical boundary layer flow simulations, $\Delta$ attains its minimum at the wall, i.e., $\min[\Delta^+]=\left.\Delta_y^+\right|_{y=0}=O(1)$. 
It then follows from Equation \eqref{eq:DCFL} that 
\begin{equation} 
\begin{split}
    dt\leq&\min_{x}\left[\frac{\nu}{u_\tau^2}\right],~~{\rm for~WRLES~and~DNS}.\\
\end{split}
\end{equation}
Invoking Equation \eqref{eq:emp}, we have
\begin{equation} 
\begin{split}
    \frac{U_\infty^2 dt}{\nu}\leq&74~{Re_{x_0}^{1/7}},~~{\rm for~WRLES~and~DNS}.\\
    \label{eq:dt-dcfl}
\end{split}
\end{equation}
For WMLES, because the wall layer is not resolved, the viscous stability condition is not a concern.
Comparing Equation \eqref{eq:dt-dcfl} with Equation \eqref{eq:dt5}, the time steps required to resolve the convective and diffusive physical time scales have the same asymptotic dependence on Reynolds number.
An implication of this result is that implicit treatment of the viscous terms typically only leads to a computational speedup by a constant factor.  
}

\section{Conclusion}
\label{sec:conclusion}

{The estimates for the grid-point and time-step requirements for DNS, WRLES, and WMLES are summarized in Table \ref{tab:summary}. 
}
These estimates depend critically on how one deploys the grid points.
Choi \& Moin assumed that $dx$, $dy$, and $dz$ scales with the Kolmogorov length scale at the wall, and they concluded $N_{\rm DNS}\sim Re_{L_x}^{2.65}$.
Here, by allowing $dx$, $dy$, and $dz$ to depend on the local Kolmogorov length scale, we arrive at $N_{\rm DNS}\sim Re_{L_x}^{2.06}$.
Hence, the use of nested grids is critical to the efficient use of the limited computational resources.
However, the more common practice is to keep $dx$ and $dz$ constant and vary $dy$ as a function of the wall-normal coordinate, \cite{kim1987turbulence,lozano2014effect, lee2015direct, ryu2019hydrodynamics, dai2019coherent} which is an inefficient use of the computational resources.
While there are other meshing techniques, \cite{kravchenko1996zonal,toosi2017anisotropic} Voronoi mesh is a promising path to nested grids. \cite{voronoi1908new,augenbaum1985construction}
One can build a nested Voronoi mesh by clustering points according to the local grid-point requirement and creating a Voronoi diagram from these scattered points, {see, Refs  \onlinecite{lehmkuhl2018large, goc2020wall} for recent applications of this technique.}

In addition to the grid-point requirements, we estimate the time-step requirements for WMLES, WRLES, and DNS.
For a fully explicit or a semi-implicit code, the time-step requirement is $N_t\sim Re_{L_x}/Re_{x_0}^{1/7}$ for DNS, $N_t\sim Re_{L_x}/Re_{x_0}^{1/7}$ for WRLES, $N_t\sim Re_{L_x}/Re_{x_0}^{6/7}$ for WMLES, as shown in Equation \eqref{eq:nt2}. 
Since parallelizing time is very difficult, this estimate provides an optimistic outlook for WMLES.
In arriving at the above estimates for the time-step requirements, we obtain two useful conclusions: first, implicit treatment of viscous terms does not significantly benefit time stepping, and second, the most limiting region in terms of time stepping is the buffer layer for DNS rather than the viscous sublayer.
Both conclusions, however, are true only if the grid is such that $dx$, $dy$, and $dz$ scale with the local Kolmogorov length scale.
In practice, $dx$ and $dy$ may be such that $dx^+/dy^+  = \mathcal{O}(10-100)$ in the wall layer. \cite{lozano2014effect,lee2015direct}
In that case, implicit treatment of the viscous term would benefit time stepping.

{According to the previous estimates (in Ref \onlinecite{choi2012grid}) and the numbers in the last column of Table \ref{tab:summary}, the saving in the computational cost by going from DNS to WRLES is nearly as significant as from WRLES to WMLES.
Because of this prior estimate, WRLES is often perceived as a cost-effective alternative to DNS.
The present estimate conveys a very different message.
According to the present estimate and the numbers in Table \ref{tab:summary}, if one uses a nested grid and a semi-implicit code, the saving in the computational cost by going from DNS to WRLES is much smaller than that by going from WRLES to WMLES, making WRLES, a computational tool that relies on sub-grid scale modeling, much less attractive.
Considering that DNS of flows at practically relevant Reynolds numbers is not going to be possible in the near future, the present estimates give a very promising outlook for WMLES.
}

\begin{table}[] 
{
\caption{\label{tab:summary}A summary of Reynolds number scalings of the present and prior grid-point requirements, time-step requirements, and the overall simulation cost. 
Here, we assume $x_0/L_x$ is Reynolds number independent (see Section \ref{sec:time_step} for detailed discussion). 
$^*$ Chapman and Choi \& Moin did not formally estimate the time-step requirements; therefore, the estimates of the overall cost are made using the conventional back-of-the-envelope calculation in Refs \onlinecite{reynolds1990potential,piomelli2002wall}, i.e., $N_t\sim N^{1/3}$.}
}
\begin{spacing}{1.25}
\begin{tabular}{l|ccc|ccc|ccc|}
\cline{2-10}

                                   & 
                                    \multicolumn{3}{c|}{
                                        \makecell[cc]{
                                                Grid-point exponent $a$,\\ where $N \sim Re_{L_x}^a$
                                        }
                                   }                                                         & \multicolumn{3}{c|}{
                                   \makecell[cc]{Time-step exponent $b$, \\ where $N_t \sim Re_{L_x}^b$}}                                                                                                                        & \multicolumn{3}{c|}{
                                   \makecell[cc]{Overall cost exponent c, \\ where $N N_t \sim Re_{L_x}^c$.}}                                                 \\ \cline{2-10} 
                                   & \multicolumn{1}{c|}{$~~~~$DNS$~~~~$} & \multicolumn{1}{c|}{$~~$WRLES$~~$} & $~~$WMLES$~~$  & \multicolumn{1}{c|}{$~~~~$DNS$~~~~$}              & \multicolumn{1}{c|}{$~~$WRLES$~~$}                & $~~$WMLES$~~$                                     & \multicolumn{1}{c|}{$~~~~$DNS$~~~~$} & \multicolumn{1}{c|}{$~~$WRLES$~~$} & $~~$WMLES$~~$     \\ \hline
\multicolumn{1}{|l|}{Chapman \cite{chapman1979computational}}      & -                                      & ${1.8}$                     & ${0.4}$ & -                                                   & -                                                   & -                                                   & -                                      & ${2.4}^*$                   & ${0.53}^*$ \\ \cline{1-1}
\multicolumn{1}{|l|}{Choi \& Moin \cite{choi2012grid}} & ${2.64}$                      & ${1.86}$                    & ${1.0}$ & -                                                   & -                                                   & -                                                   & ${3.52}^*$                    & ${2.48}^*$                  & ${1.33}^*$ \\ \cline{1-1}
\multicolumn{1}{|l|}{Present}      & ${2.05}$                      & ${1.86}$                    & ${1.0}$ & ${6/7}$ & ${6/7}$ & ${1/7}$ & ${2.91}$                      & ${2.72}$                    & ${1.14}$   \\ \hline
\end{tabular}
\end{spacing}
\end{table}

\section*{Acknowledgement}
XY acknowledges financial support from the Elliott Group and the Office of Naval Research.
KG acknowledges support from the National Defense Science and Engineering Graduate Fellowship and the Stanford Graduate Fellowship.
XY and KG thank Parviz Moin, Ugo Piomelli, Adrian Lozano-Duran and Vishal Jariwala for their useful comments.

\section*{DATA AVAILABILITY}
The data that support the findings of this study are available from the corresponding author upon reasonable request.

\appendix
\section{Nomenclature}

{In this appendix, we list the variables used in our manuscript.}

\begin{itemize}
\setlength\itemsep{0em}
    \item[] $C$ ~~ Constant
    \item[] $c_f$ ~~ friction coefficient
    \item[] $dt$ ~~ time step size
    \item[] $L$ ~~ length
    \item[] $N$ ~~ total number of grid points
    \item[] $n$ ~~ constant describing WRLES resolution
    \item[] $N_t$ ~~ number of time steps
    \item[] $Re$ ~~ Reynolds number
    \item[] $T$ ~~ time scale
    \item[] $U$ ~~ mean streamwise velocity
    \item[] $U_\infty$ ~~ freestream velocity
    \item[] $u_\tau$ ~~ friction velocity
    \item[] $u'$ ~~ streamwise velocity fluctuation
    \item[] $v'$ ~~ wall-normal velocity fluctuation
    \item[] $x$ ~~ streamwise coordinate
    \item[] $x_0$ ~~ inlet location
    \item[] $y$ ~~ wall-normal coordinate
    \item[] $z$ ~~ spanwise coordinate
\end{itemize}

\noindent
Greek letters
\begin{itemize}
\setlength\itemsep{0em}
    \item[] $\delta$ ~~ boundary layer thickness
    \item[] $\Delta$ ~~ grid spacing
    \item[] $\epsilon$ ~~ turbulent dissipation
    \item[] $\kappa$ ~~ von Karman constant
    \item[] $\eta$ ~~ Kolmogorov length scale
    \item[] $\nu$ ~~ kinematic viscosity
\end{itemize}

\noindent
Subscripts
\begin{itemize}
\setlength\itemsep{0em}
    \item[] $w$ ~~ wall
    \item[] $wr$ ~~ wall resolved
    \item[] $eto$ ~~ eddy turnover
    \item[] $ft$ ~~ flow through
\end{itemize}

\section*{References}
\bibliographystyle{ieeetr}
\bibliography{WMLES,RoChannel}



\end{document}